\documentclass[pra,aps,twocolumn]{revtex4}
\usepackage{graphicx}

\def \etal{{\em et al.}}
\def \g{\downarrow}
\def \e{\uparrow}

\def \>{\rangle}
\def \<{\langle}

\begin{document}

\title{Quantum Entanglement and Teleportation of\\
Quantum-Dot States in Microcavities}

\author{A. Miranowicz}
\affiliation{SORST-JST, Honmachi, Kawaguchi, Saitama 331-0012,
Japan} \affiliation{Graduate School of Engineering Science, Osaka
University, Toyonaka, Osaka 560-8531, Japan}
\affiliation{Institute of Physics, Adam Mickiewicz University,
61-614 Pozna\'n, Poland}

\author{\c{S}. K. \"Ozdemir}
\affiliation{SORST-JST,~Honmachi, Kawaguchi, Saitama 331-0012, Japan,\\
CREST-JST, Honmachi, Kawaguchi, Saitama 331-0012, Japan and\\
Graduate School of Engineering Science, Osaka
University, Toyonaka, Osaka 560-8531, Japan}

\author{Yu-xi Liu}
\affiliation{Frontier Research System,  Institute of Physical and
Chemical Research (RIKEN), Wako-shi 351-0198, Japan}

\author{G. Chimczak}
\affiliation{Institute of Physics, Adam Mickiewicz University,
61-614 Pozna\'n, Poland}

\author{M. Koashi}
\author{N. Imoto}
\affiliation{SORST-JST, Honmachi, Kawaguchi, Saitama 331-0012, Japan,\\
CREST-JST, Honmachi, Kawaguchi, Saitama 331-0012, Japan and\\
Graduate School of Engineering Science, Osaka
University, Toyonaka, Osaka 560-8531, Japan}

\begin{abstract}
Generation and control of quantum entanglement are studied in an
equivalent-neighbor system of spatially-separated semiconductor
quantum dots coupled by a single-mode cavity field. Generation of
genuinely multipartite entanglement of qubit states realized by
conduction-band electron-spin states in quantum dots is discussed.
A protocol for quantum teleportation of electron-spin states via
cavity decay is briefly described.
\end{abstract}

\pagenumbering{arabic} \maketitle

\section{Introduction}

Among various proposals of scalable quantum computers
\cite{book1}, there has been an increasing interest in
quantum-information processing (QIP) with quantum dots (QDs)
\cite{book2} since the seminal work of Loss and DiVincenzo
\cite{Loss98}. Recently, the interest has further been stimulated
by experiments demonstrating the viability of the coherent
manipulation of charge states in a single QD \cite{spin1} and a
double QD \cite{spin2}.

Quantum bits (qubits) can be implemented in nanostructures in
various ways including electron-spin states (as, e.g., discussed
in the next Section), excitonic states \cite{Loss98} or
nuclear-spin states \cite{Leuenberger,Hirayama,Ozdemir}. The main
advantage of the spin qubits is the decoherence times that are a
few orders of magnitude longer than the other relevant time
scales. However, in a practical implementation of the QD-based
quantum computer, one should also be able to (i) quickly induce
and control the long-distance couplings between selectively chosen
QDs and to (ii) scale such computer to hundreds of qubits, which
seemingly require fabrication of high-quality, regularly spaced,
uniform semiconductor QDs. A scheme of Imamo\v{g}lu
\etal~\cite{Imamoglu} offers a possible way to overcome the
above-mentioned problems by placing the QDs in a microcavity and
illuminating them by laser beams. The long-distance QD
interactions are mediated by a single-mode cavity field, and their
control is realized by addressing selectively the chosen QDs by
laser beams. The possible irregularity of QD structures can be
overcome by choosing the proper frequencies and intensities of the
laser fields. It is worth noting that a crucial condition for a
realization of QIP in such models is a strong coupling of a single
QD to a single mode of microcavity (or nanocavity) of a high
quality factor (high-Q). Quite recently such random couplings
\cite{Reithmaier} or even deterministic couplings \cite{Badolato}
have been observed experimentally.

Quantum entanglement is a key resource for QIP \cite{book1}. Here,
we extend our former results on bipartite entanglement generation
in QDs \cite{Miranowicz,Liu02a,Liu02b,Liu04} by analysing also
generation of multipartite entanglement in the QD systems. We also
suggest a protocol for teleportation of QD spin states between
distant cavities via their decay. It is a generalization of the
scheme for teleportation of atomic-qubit states via cavity decay
\cite{Bose,Chimczak1,Chimczak2,Chimczak3}.

It is worth stressing that the present analysis is focused on
entanglement between electron spins of typical QDs, i.e., of the
size smaller than the Bohr radius. In Refs.
\cite{Liu02a,Liu02b,Liu04,Liu03}, we studied quantum entanglement
of excitons in systems of QDs of the size larger than the Bohr
radius of an exciton in bulk semiconductor but smaller than the
relevant optical wavelengths. In particular, we described
realizations of the entangled webs of QD excitons with symmetric
sharing of entanglement, where each QD is equally entangled to all
others. The decoherence of the generated maximally entangled
states was studied in greater detail. In particular, we predicted
decoherence times as a function of the size of the GaAs and CdS
large QDs \cite{Liu03}.

\section{A model of quantum dots in a microcavity}

We are interested in quantum-information properties of
electron-spin states of semiconductor QDs within the model of
Imamo\v{g}lu \etal~\cite{Imamoglu}, which can be described as
follows: The QDs are placed on a microdisk, put into a microcavity
tuned to frequency $\omega _{{\rm cav}}$, and illuminated
selectively by laser fields of frequencies $\omega _{n}^{L}$,
where $n$ labels the QDs. Each of $N$ QDs with a single electron
in the conduction band is modeled by a three-level atom as shown
in figure 1. The total Hamiltonian for the three-level QDs
interacting with quantized fields reads as follows
\cite{Miranowicz}:
\begin{eqnarray} \label{N01}
\hat{H}&=&\hat{H}_{\rm QD}+\hat{H}_{\rm fields}+\hat{H}_{\rm int},
\\
\hat{H}_{\rm QD}&=&\sum_{n=1}^N \Big[{\cal E}_{n}^{\g}\hat{\sigma}
_{n}^{\g\g}+{\cal E}_{n}^{\e}\hat{\sigma} _{n}^{\e\e} +{\cal
E}_{n}^{v}\hat{\sigma} _{n}^{vv}\Big],
  \nonumber \\
\hat{H}_{\rm fields}&=&\hbar \omega _{{\rm cav}}\hat{a}_{{\rm
cav}}^{\dag }\hat{a}_{{\rm cav} }+\sum_{n=1}^N\hbar \omega
_{n}^{L}(\hat{a}_{n}^{L})^{\dag }\hat{a}_{n}^{L},
  \nonumber \\
\hat{H}_{\rm int} &=&\sum_{n=1}^N \Big[\hbar
g_{n}^{v\g}\hat{a}_{n}^{L}\hat{\sigma} _{n}^{\g v}+\hbar
g_{n}^{v\e}\hat{a}_{{\rm cav}}\hat{\sigma} _{n}^{\e v}+{\rm h.c.}
\Big],
  \nonumber
\end{eqnarray}
where $\hat{H}_{\rm QD}$ and $\hat{H}_{\rm fields}$ are the free
Hamiltonians of the QDs and the fields, respectively;
$\hat{H}_{\rm int}$ is the interaction Hamiltonian; $\hat{a}_{\rm
cav}$ and $\hat{a}^{\dag}_{\rm cav}$ are the annihilation and
creation operators of the cavity mode, respectively;
$\hat{a}^{L}_{n}$ and $(\hat{a}^{L}_{n})^{\dag}$ are the
corresponding operators for the laser modes; $\hat{\sigma}
_{n}^{xy}=|x\rangle _{nn}\langle y|$ is the $n$th QD operator;
${\cal E}_{n}^{(x)}$ is the energy of level $|x\rangle_{n}$
($x=\g,\e,v$); the $n$th QD levels $|\g\rangle _{n}$ and
$|v\rangle _{n}$ are coupled by dipole interactions with a
strength of $g_{n}^{v\g}$; analogously, $g_{n}^{v\e}$ is the
coupling strength between levels $|\e\rangle _{n}$ and $ |v\rangle
_{n}$. There is no direct coupling between levels $|\g\rangle _{n}
$ and $|\e\rangle _{m}$ in either the same ($n=m$) or different
QDs ($n\neq m$). The Hamiltonian (\ref{N01}) simply generalizes,
to $N$ QDs and $N+1$ fields, the standard quantum-optical models
of a three-level system interacting with two radiation modes (see
e.g. \cite{3plus2}).
   \begin{figure}
   \includegraphics[height=4cm]{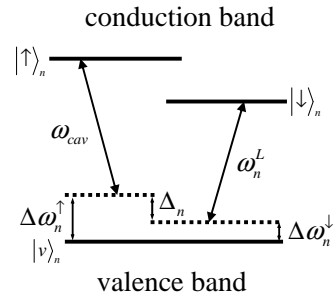}
   \vspace{-1mm}
   \caption{
Effective energy levels of the $n$th QD. Key: $|v\rangle _{n}$ --
the effective valence-band state of energy ${\cal E} _{n}^{v}$;
$|\e\rangle _{n}$ ($|\g\rangle _{n}$) -- spin up (spin down) state
of the conduction-band electron of energy ${\cal E} _{n}^{\e}$
(${\cal E} _{n}^{\g}$); detunings are defined by $\hbar\Delta
\omega _{n}^{\e}={\cal E}_{n}^{\e}-{\cal E} _{n}^{v}-\hbar\omega
_{{\rm cav}}$, $\hbar\Delta \omega _{n}^{\g}={\cal E}
_{n}^{\g}-{\cal E} _{n}^{v}-\hbar\omega _{n}^{L}$ and $\Delta
_{n}=\Delta \omega _{n}^{\e}-\Delta \omega _{n}^{\g}$.}
\end{figure}

By applying an adiabatic elimination method, Imamo\v{g}lu
\etal~\cite{Imamoglu} derived the effective interaction
Hamiltonian
\begin{eqnarray}
\hat{H}_{{\rm eff}}&=&\!\frac{\hbar}{2}\sum_{n\neq m}\kappa
_{nm}(t)\Big({\hat{\sigma}}
_{n}^{+}{\hat{\sigma}}_{m}^{-}e^{i(\Delta _{n}-\Delta _{m})t}
+{\rm h.c.} \Big) \label{N02}
\end{eqnarray}
describing the evolution of the conduction-band spins of $N$ QDs
coupled by a microcavity field. This Hamiltonian is given in terms
of the Pauli spin creation ${\hat{\sigma}}_{n}^{+}$ and
annihilation ${\hat{\sigma}}_{n}^{-}$  operators acting on the
conduction-band spin states of the $n$th QD. The effective
strength of two-QD coupling between the spins of the $n$th and
$m$th QDs is given by $\kappa _{nm}(t)= g_n(t) g_m(t)/\Delta
_{n}$, where the effective single-QD coupling of the $n$th spin to
the cavity field is $g_n(t)= g_{n}^{v\g} g_{n}^{v\e}
|E^{L}_n(t)|/\Delta \omega _{n}$ with $\Delta \omega _{n}$ being
the harmonic mean of $\Delta \omega _{n}^{\e}$ and $\Delta \omega
_{n}^{\g}$. For simplicity, the laser fields are assumed to be
strong and treated classically as described by the complex
amplitudes $E^{L}_n(t)$. The Hamiltonian (\ref{N02}) was derived
by applying adiabatic eliminations of the valence-band states
$|v\rangle _{n}$ and cavity mode $\hat{a}_{{\rm cav}},$ which are
valid under the assumptions of negligible coupling strength,
cavity decay rate, and thermal fluctuations in comparison to
$\hbar\Delta _{n}$, $\hbar\Delta \omega _{n}^{\g}$, $\hbar\Delta
\omega _{n}^{\e}$ and the energy difference ${\cal
E}_{n}^{\e}-{\cal E} _{n}^{\g}$ (see figure 1). Moreover, the
valence-band levels $|v\rangle _{n} $ were assumed to be far off
resonance. Although the Hamiltonian (\ref{N02}) describes
apparently direct spin-spin interactions, the real physical
picture is different as the quantum-QD spins are coupled only
indirectly via the cavity field as described by Hamiltonian
(\ref{N01}).

As discussed in Ref. \cite{Miranowicz}, Eq. (\ref{N02}) can be
reduced to the effective equivalent-neighbor $N$-QD Hamiltonian
\begin{eqnarray}
\hat{H}_{\rm eff}=\frac{\hbar\kappa}{2} \sum_{n\neq m}\left(
{\hat{\sigma}}_n^{+}{\hat{\sigma}}_m^{-}+
{\hat{\sigma}}_n^{-}{\hat{\sigma}}_m^{+}\right) \label{N03}
\end{eqnarray}
even for nonidentical QDs, which can be achieved by adjusting the
laser-field frequencies $\omega _{n}^{L}$ to get the same detuning
$\Delta _{n}=\,$const, and by choosing the proper laser
intensities $|E^{L}_n|^2$ to obtain the effective coupling
constants of $g_{n}(t)=g$ or, equivalently, $\kappa _{nm}(t)\equiv
\kappa$ independent of subscripts $n$ and $m$. Thus, in a special
case, the model of Imamo\v{g}lu \etal~offers a physical
realization of the equivalent-neighbor-QD interactions, where each
QD interacts with all others with the same strength regardless of
their positions or differences in their energetic levels. Note
that a system of equivalent-neighbor interactions of up to only
four particles can be realized by placing particles in a symmetric
geometric configuration. But in the discussed equivalent-neighbor
model, the number of QDs can practically be scaled up to $N\sim
100$.

In order to write compactly a solution of the model, it is
convenient to introduce the (unnormalized) totally symmetric state
\begin{eqnarray}
  |\Phi_{m,n}\> &=&  \{|\e\rangle ^{\otimes m}|\g\rangle ^{\otimes n}\}
\label{N04}
\end{eqnarray}
as a sum of all $(n+m)$-QD states with $m$ spins up and $n$ spins
down. In particular, state (\ref{N04}) for $m=1,n=1$ corresponds
to a Bell state (one of the triplet states), $|B\> =
\frac{1}{\sqrt{2}}|\Phi_{1,1}\> = \frac{1}{\sqrt{2}} (|\e\g\rangle
+|\g\e\rangle)$, while for $m=1,n=2$ (and analogously for
$m=2,n=1$) corresponds to the $W$ state, i.e.:
\begin{eqnarray}
  |W\> &=& \frac{ |\Phi_{1,2}\>} {\sqrt{3}}  = \frac{1}{\sqrt{3}}
  (|\e\g\g\>+|\g\e\g\>+|\g\g\e\>).
\label{N05}
\end{eqnarray}
Let us assume that the initial state describing a system of $M$
($M=0,\ldots ,N$) QDs with spin up (of a single conduction-band
electron) and $(N-M)$ QDs with spin down is $|\psi^{NM}(0)\rangle
=|\e\rangle ^{\otimes M}|\g\rangle
^{\otimes(N-M)}=|\Phi_{M,0}\>|\Phi_{0,N-M}\>$. Then the solution
of the Schr\"odinger equation of motion for Hamiltonian
(\ref{N03}) is given by \cite{Miranowicz}:
\begin{eqnarray}
\label{N06} |\psi^{NM} (t)\rangle =\sum_{m=0}^{M^{\prime }}
\gamma_{m}^{NM}(t) |\Phi_{M-m,m}\> |\Phi_{m,N-M-m}\>.
\end{eqnarray}
The time-dependent superposition coefficients are given by
\begin{eqnarray}
\gamma_{m}^{NM}(t)&=&\sum_{n=0}^{M^{\prime }}b_{nm}^{NM}
e^{i[n(N+1-n) -M(N-M)]\kappa t},
\label{N07} \\
b_{nm}^{NM}&=&\sum_{k=0}^{m} (-1)^{k} \frac{C^m_k}{C^{N-2k}_{M-k}}
\Big(C^{N+1-2k}_{n-k}-2C^{N-2k}_{n-k-1} \Big), \label{N08}
\end{eqnarray}
where $M^{\prime }=\min (M,N-M)$ and $C^{x}_{y}$ are binomial
coefficients. For $N$-QD systems initially in $|\psi^{N1}(0)\> =
{|\e\>} {|\g\>^{\otimes(N-1)}}$, solution (\ref{N06}) simplifies
to:
\begin{equation}
  |\psi^{N1}(t) \rangle = \gamma^{N1}_{0}(t) |\e\>|\g\>^{\otimes(N-1)}
  +\gamma^{N1}_{1}(t) |\g\>|\Phi_{1,N-2}\>.
\label{N09}
\end{equation}
We will apply and analyze quantum properties of these solutions in
the next section.

\section{Entanglement of electron-spin states}

Here, we analyze generation of quantum entanglement of
electron-spin states of semiconductor QDs within the discussed
equivalent-neighbor model. We will describe various kinds of
entanglement as depicted schematically in figure 2 including pure-
and mixed-state bipartite entanglement, as well as pure-state
genuine multipartite entanglement.

In the analysis of the pure-state bipartite entanglement, it is
useful to decompose the solution (\ref{N06}) as follows
\begin{equation}
|\psi^{NM} (t)\rangle =\sum_{m=0}^{M^{\prime
}}\sqrt{P_{m}^{NM}(t)}|\phi _{m}(t)\rangle _{A}|\varphi
_{m}(t)\rangle _{B}  \label{N10}
\end{equation}
in terms of the orthonormal-basis states $|\phi _{m}(t) \rangle
_{A}$ and $|\varphi _{m}(t)\rangle_{B}$ for subsystems $A$ and
$B$, respectively, and the Schmidt coefficients given by
\begin{equation}
P_{m}^{NM}(t)=C^{M}_{m} C^{N-M}_{m} |\gamma_{m}^{NM}(t)|^{2}.
\label{N11}
\end{equation}
Eq. (\ref{N11}) directly enables us calculation of the von Neumann
entropy
\begin{eqnarray}
E^{NM}(t) &=&-\sum_{m=0}^{M'}P_{m}^{NM}(t)\log _{2}P_{m}^{NM}(t),
\label{N12}
\end{eqnarray}
as the Shannon entropy of the Schmidt coefficients. This formula
determines bipartite entanglement between the QDs initially with
spin up (subsystem $A$) and the remaining QDs (subsystem $B$), as
shown in figure 2(a). It is easy to see that it holds the
following symmetry of the Schmidt coefficients
$P_{m}^{N,M}(t)=P_{m}^{N,N-M}(t)$, which implies that the
entanglements for systems with $M$ and $(N-M)$ QDs initially with
spin up are the same for any evolution times.

   \begin{figure}
   \hspace*{0mm}\includegraphics[height=6cm]{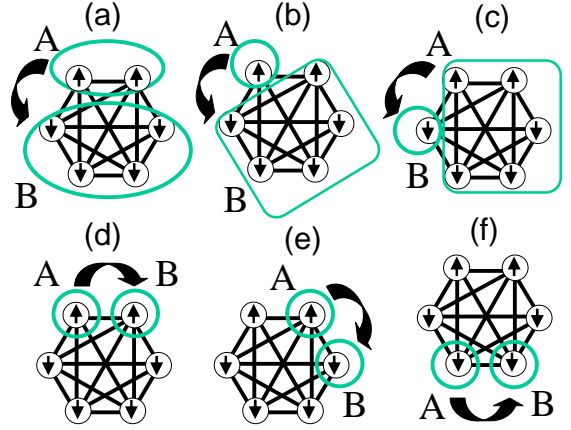}
   \vspace{-4mm}
   \caption{
Schematic representation of various bipartite entanglements
between subsystems $A$ and $B$ in a system of $N$
equivalent-neighbor QDs: (a) pure-state entanglement between $M$
QDs initially with spin up and the remaining $(N-M)$ QDs with spin
down as described by the von Neumann entropy $E^{NM}(t)$; (b,c)
pure-state entanglement between a given QD (initially either with
spin up or down) and the rest of the system being described by
tangle $\tau^{NM}_{\e\tilde\e}(t)$ or $\tau^{NM}_{\g\tilde\g}(t)$
respectively; (d,e,f) three types of mixed-state pairwise
entanglement described by concurrences $C^{NM}_{\e\e}(t)$,
$C^{NM}_{\e \g}(t)$, and $C^{NM}_{\g\g}(t)$, respectively.
Analysis of bipartite entanglement shown in figures (b)--(f)
enables extraction of information about intrinsic multipartite
entanglement.}
\end{figure}

In Ref. \cite{Miranowicz}, we addressed the question of generation
of maximum entanglement between two subsystems $A$ and $B$
consisting of $M$ and $(N-M)$ QDs, respectively. The simplest
nontrivial evolution of our system occurs for two QDs with one of
them initially with spin up given by
\begin{eqnarray}
 |\psi^{21}(t)\> &=&
 \gamma^{21}_{0}(t) |\e\g\>+ \gamma^{21}_{1}(t) |\g\e\>
\nonumber\\
 &=& \cos(\kappa t) |\e\g\>-i \sin(\kappa t) |\g\e\>,
 \label{N13}
\end{eqnarray}
as a special case of (\ref{N09}). The state periodically evolves
into Bell-like states $|\psi ^{21}(t^{\prime })\rangle =
\frac{1}{\sqrt{2}} (|\e\g\rangle \pm i|\g\e\rangle)$ at times
$\kappa t^{\prime }=(2n+1)\frac \pi 4$ ($n=0,1,\dots$). In
general, from the Bennett {\em et al.} theorem \cite{Bennett96}
follows that the maximally entangled states has the entanglement
of $\log _{2}\{\min (M,N-M)+1\}$ ebits. For our systems under
interactions described by (\ref{N03}) with a single spin up
($M=1$), the Schmidt coefficients (\ref{N11}) reduce to
$P_{1}^{N1}(t)=4(N-1) N^{-2}\sin ^{2}(\frac{N}{2}\kappa t)$ and
$P_{0}^{N1}(t)=1-P_{1}^{N1}(t)$, which enable a direct calculation
of the entanglement $E^{N1}(t)$ and its maximum values from Eq.
(\ref{N12}). One can find \cite{Miranowicz} that the maxima can be
observed at the times $\kappa t^{\prime }=\pm {{\frac{2}{N}}}{\rm
arccsc}({{\frac{2}{N}}}\sqrt{2(N-1)})+nT$ and $\kappa t^{\prime
\prime }=\frac{\pi }{N}+nT$ with $n=0,1,\cdots$ multiples of the
period $T\equiv T^{N1}=\frac{2\pi}{N}$. Then it is easy to prove
that the maximally entangled state (EPR state) with
${E}^{N1}(t^{\prime })=1$ ebit, can be achieved at the evolution
times $t^{\prime}$ for $N\leq 6$ only. For $N>6$, any real
solution for $t^{\prime}$ does not exist, and the entanglement
reaches its maximum at the evolution times $t^{\prime \prime }$
but ${E}^{N1}(t'')=\max_t{E}^{N1}(t)$ is less than one ebit, thus
the EPR state cannot be generated exactly in the systems having
more than six QDs, where one of them has initially spin opposite
to the spin of the other QDs. A numerical analysis shows that if
the number $M$ of initial spins up is $1<M<N-1$ of any number $N$
of QDs then our system will not arrive at the exact EPR states
either. A sudden decrease of entanglement is observed on
increasing the total number of QDs in a system with a fixed number
of QDs with spin up. Nevertheless, very good approximations of the
EPR states with almost $\log _{2}([N/2]+1)$ ebits of entanglement
can be achieved periodically if the system has half (or almost
half) spins up, $M=[N/2]$ \cite{Miranowicz}.

To analyze a genuine multipartite entanglement, we apply the
approach proposed by Coffman, Kundu and Wootters (CKW) \cite{CKW}
via the so-called tangles and monogamy inequality. Let us define
the tangle $\tau(\hat{\rho}_{n\tilde{n}})$ as the entanglement
measure between the $n$th qubit and all the remaining ones
(denoted by $\tilde{n}$), which corresponds to entanglement
schematically depicted in figures 2(b,c). The tangle for arbitrary
mixed state $\hat{\rho}=\hat{\rho}_{n\tilde{n}}$ of the $2\times
d$ system is defined as the convex roof \cite{Osborne}:
\begin{eqnarray}
\tau (\hat{\rho}_{n\tilde{n}})=\inf_{\{p_{i},|\psi _{i}\rangle
\}}\sum_{i}p_{i}S[\text{tr}_{\tilde{n}}(|\psi _{i}\rangle \langle
\psi _{i}|)] \label{N14}
\end{eqnarray}
of the single-qubit linear entropy
\begin{eqnarray}
  S(\hat{\rho})=2[1-\text{tr(}\hat{\rho}^{2})]=4\det
(\hat{\rho}), \label{N15}
\end{eqnarray}
where the infimum is taken over all pure-state decompositions
$\{p_{i},|\psi _{i}\rangle \}$ of $\hat{\rho}=p_{i}|\psi
_{i}\rangle \langle \psi _{i}|.$ For pure states, as analyzed in
our paper, the tangle is simply defined as $\tau
(\hat{\rho}_{n\tilde{n}})=S[\text{tr}_{\tilde{n}}
(\hat{\rho}_{n\tilde{n}})].$

A pairwise entanglement between any two QDs, as depicted
schematically in figures 2(d,e,f), can be described by the
concurrence, defined by \cite{Wootters}
\begin{equation}
C({\hat \rho })=\max \{0,2\max_{i}\lambda
_{i}-\sum_{i=1}^{4}\lambda _{i}\} \label{N16}
\end{equation}
for a reduced two-qubit mixed state $\hat \rho$. In definition
(\ref{N16}), $\lambda _{i}$ are the square roots of the
eigenvalues of the matrix ${\hat \rho }({\hat \sigma }_{y}\otimes
{\hat \sigma }_{y}){\hat \rho }^{\ast }({ \hat \sigma }_{y}\otimes
{\hat \sigma }_{y})$, where ${\hat \sigma }_{y}$ is the Pauli spin
matrix and the asterisk denotes complex conjugation. The
concurrence is related to the entanglement of formation,
$E_{F}({\hat \rho })$, defined as the minimum mean entanglement of
an ensemble of pure states $|\psi _{i}\rangle $ that represents
$\hat \rho$ \cite{Bennett96a}: $ E_{F}({\hat \rho
})=\min_{\{p_{i},|\psi _{i}\rangle \}}\sum_{i}p_{i}E(|\psi
_{i}\rangle \langle \psi _{i}|),$ where $\hat \rho
=\sum_{i}p_{i}|\psi _{i}\rangle \langle \psi _{i}|$ and $E(|\psi
_{i}\rangle \langle \psi _{i}|)$ is the entropy of entanglement of
pure state $|\psi _{i}\rangle $ defined by the von Neumann
entropy. As shown by Wootters \cite{Wootters}, the entanglement of
formation for two qubits in an arbitrary mixed state ${\hat \rho
}$ can explicitly be given in terms of the concurrence as follows
$ E_{F}({\hat \rho })= H(\textstyle {\frac{1}{2}} [1+\sqrt{1-C
({\hat \rho })^{2}}]),$ where $H(x)$ is the Shannon binary
entropy. Note that $E_{F}(\hat \rho)$ and $C(\hat \rho)$ are
monotonic functions of one another and both range from 0 (for a
separable state) to 1 (for a maximally entangled state).

CKW conjectured that the bipartite entanglement of multipartite
qubit states satisfies the following monogamy inequality
\cite{CKW}
\begin{eqnarray}
 \Delta (\hat{\rho}_{n})=\tau (\hat{\rho}_{n\tilde{n}})-\sum_{m
(m\neq n)} C^{2}(\hat{\rho}_{nm})\geq 0 \label{N17}
\end{eqnarray}
as quantified by the squared concurrence $C^{2}(\hat{\rho}_{nm})$
and the tangle $\tau (\hat{\rho}_{n\tilde{n}})$. The conjecture
was first proved by CKW for 3 qubits in pure state \cite{CKW}, but
a general proof for arbitrary number of qubits in arbitrary states
has been given only recently by Osborne and Verstraete
\cite{Osborne}. We use the quantity $\Delta (\hat{\rho}_{n})$,
which is called  the {\em residual tangle}, as a criterion for
genuine (or intrinsic) multipartite entanglement of qubit states,
i.e. for correlations not stored in two-qubit entanglement. In
particular, the residual tangle for $N=3$ qubits is referred as
the 3-tangle and describes genuine 3-partite (3-way) entanglement.
In general, for $N>3$ qubits, $\Delta (\hat{\rho}_{n})$ contains
information about not solely $N$-way entanglement, but rather all
kinds of $3,4,...,N$-partite entanglements.

   \begin{figure}
   \includegraphics[height=8cm]{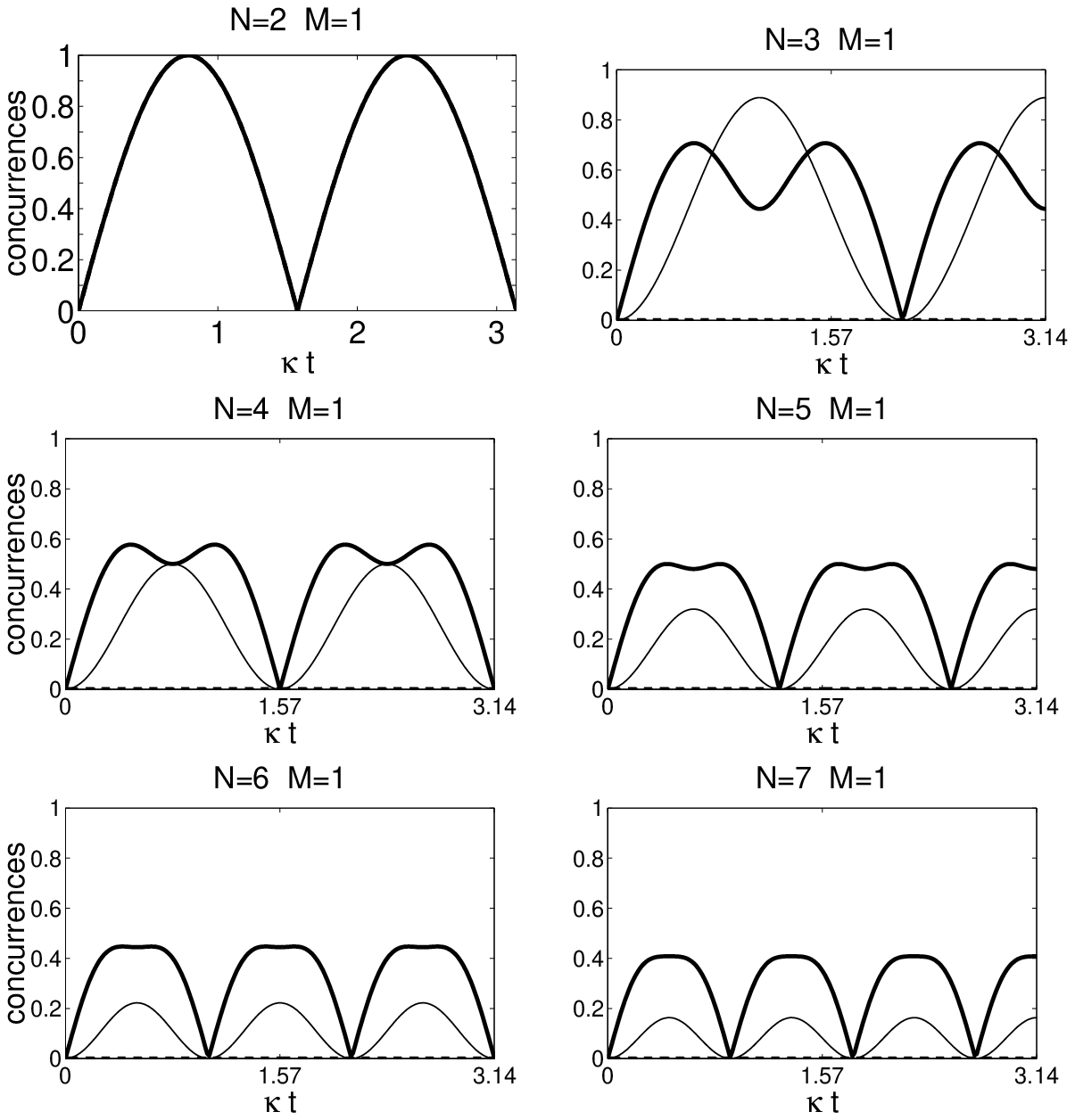}
   \vspace{-4mm}
   \caption{
Evolution of concurrence for systems of $N$ QDs with only a single
QD initially with spin up ($M=1$): $C^{N1}_{\e \g}(t)$ (thick
curves) and $C^{N1}_{\g\g}(t)$ (thin curves). Note that the
residual tangles are vanishing, $\Delta^{N1}_{\e}(t)
=\Delta^{N1}_{\g}(t)=0$.}
\end{figure}

   \begin{figure}
   \includegraphics[height=6cm]{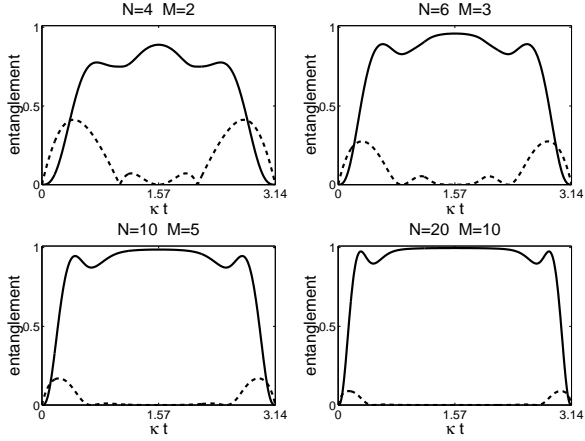}
   \vspace{-4mm}
   \caption{
Evolution of the residual tangles $\Delta^{NM}_{\e}(t)
=\Delta^{NM}_{\g}(t)$ (solid curves) and the concurrence
$C^{NM}_{\e \g}(t)$ (broken curves) for systems of $N$ QDs with
half of them with spin up. Note that $C^{NM}_{\g\g}(t)=\allowbreak
C^{NM}_{\e\e}(t)=0$.}
\end{figure}

   \begin{figure}
   \includegraphics[height=3.2cm]{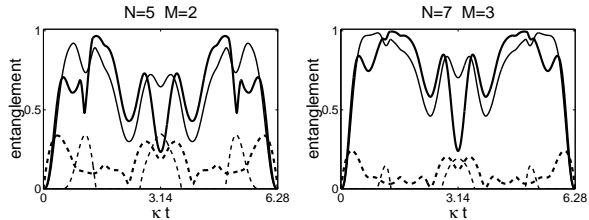}
   \vspace{-4mm}
   \caption{
Evolution of the residual tangles $\Delta^{NM}_{\e}(t)$ (thick
solid curves) and $\Delta^{NM}_{\g}(t)$ (thin solid) as well the
concurrences $C^{NM}_{\e \g}(t)$ (thick broken curves) and
$C^{NM}_{\g \g}(t)$ (thin broken) for systems of $N=2M+1$ QDs. For
clarity we omit curves for $C^{NM}_{\e \e}(t)$.}
\end{figure}

As the main new result of this paper, we analyze a possibility to
generate an intrinsic multipartite entanglement in our model as
being described by the CKW inequality. For our equivalent-neighbor
system, there are only three different types of evolution of the
concurrence depending on the choice of a pair of QDs as described
by the reduced density matrices: (i) $\hat \rho^{NM}_{\e \e}={\rm
tr}_{3,4,...,N}( \hat \rho^{NM})$ for both QDs initially with spin
up corresponding, in particular, to two qubits shown in figure
2(d), (ii) $\hat \rho^{NM}_{\e \g}={\rm
tr}_{1,2,...,M-1,M+2,...,N}(\hat \rho^{NM})$ for a pair of QDs
when initially one of them has spin up and the other has spin down
(see figure 2(e)), and (iii) $\hat \rho^{NM}_{\g\g}={\rm
tr}_{1,2,...,N-2}( \hat \rho^{NM})$ for both QDs initially with
spin down (see figure 2(f)), where $\hat \rho^{NM}= |\psi^{NM}\>
\langle\psi^{NM}|$. By properly grouping terms of Eq. (\ref{N06})
and applying partial trace, we find the following general
solutions for the three kinds of the reduced density matrices:
\begin{eqnarray}
\hat{\rho}_{\uparrow \downarrow }^{NM}(t) =\alpha _{\uparrow
\downarrow 0}(t)|\g\g\rangle \langle \g\g|+\alpha _{\uparrow
\downarrow 1}(t)|\e\e\rangle \langle \e\e|
\nonumber \\
\quad\quad +\sum_{m=0}^{M^{\prime
}-1}C_{m}^{M-1}C_{m}^{N-M-1}|\beta _{m}(t)\rangle \langle \beta
_{m}(t)|, \label{N18}
\end{eqnarray}
where $(j=0,1)$
\begin{eqnarray}
  \alpha _{\uparrow \downarrow j}(t)&=&\sum_{m=0}^{M^{\prime }}|\gamma
_{m}^{NM}(t)|^{2}C_{m-1+j}^{M-1}C_{m-j}^{N-M-1},
 \label{N19} \\
|\beta _{m}(t)\rangle &=&\gamma _{m}^{NM}(t)|\e\g\rangle +\gamma
_{m+1}^{NM}(t)|\g\e\rangle, \label{N20}
\end{eqnarray}
and for $k=\uparrow \uparrow ,\downarrow \downarrow $:
\begin{eqnarray}
\hat{\rho}_{k}^{NM}(t)&=&\alpha _{k0}(t)|\g\g\rangle \langle
\g\g|+2\alpha _{k1}(t)|B\rangle \langle B|
\nonumber \\
&&+\alpha _{k2}(t)|\e\e\rangle \langle \e\e|, \label{N21}
\end{eqnarray}
where $(j=0,1,2)$
\begin{eqnarray}
\alpha _{\uparrow \uparrow j}(t) &=&\sum_{m=0}^{M^{\prime
}}|\gamma _{m}^{NM}(t)|^{2}C_{m}^{N-M}C_{m-2+j}^{M-2}\;,
\nonumber \\
 \alpha
_{\downarrow \downarrow j}(t) &=&\sum_{m=0}^{M^{\prime }}|\gamma
_{m}^{NM}(t)|^{2}C_{m}^{M}C_{m-j}^{N-M-2}. \label{N22}
\end{eqnarray}
For simplicity, we dropped superscripts $NM$ in $\alpha _{kj}$ and
$|\beta _{m}(t)\rangle$. In the following we use the following
shorthand notation: $C^{NM}_{\e\e}(t)\equiv C(\hat
\rho^{NM}_{\e\e}(t))$, $\tau^{NM}_{\e\tilde\e}(t)\equiv \tau(\hat
\rho^{NM}_{\e\tilde\e}(t))$, $\Delta^{NM}_{\e}(t)\equiv
\Delta(\hat \rho^{NM}_{\e}(t))$, etc.

In figures 3-5, we present examples of the concurrence evolution
according to solutions for the reduced density matrices, given by
(\ref{N18}) and (\ref{N21}), assuming $M=1$ and $M=[N/2]$. Note
that $C^{N1}_{\e\e}(t)=0$ as the system has only single spin up
during the whole evolution. Analogously, one can find that
$C^{2M,M}_{\e\e}(t)=C^{2M,M}_{\g\g}(t)=0$ for any $M$. By
analyzing the figures, one can clearly see that the maximum of
concurrences decreases but the maximum of residual tangles
increases with increasing number $N$ of QDs for $M>1$.

In our equivalent-neighbor model, we find only two kinds of
tangles $\tau_{k\tilde k}^{NM}$ corresponding to a QD initially
with spin up ($k=\e$) or spin down ($k=\g$) as given by
\begin{eqnarray}
\tau_{k\tilde k}^{NM}(t) &=& 4\alpha_k(t)[1-\alpha_k(t)],
\label{N23}
\end{eqnarray}
which corresponds to the reduced density matrices $ \rho_{k\tilde
k}^{NM}(t) = \alpha_k(t) |\g\>\<\g| + [1-\alpha_k(t)] |\e\>\<\e|$,
where coefficients
$\alpha_k(t)=\alpha_{k0}(t)+\alpha_{k1}(t)=1-\alpha_{k1}(t)-\alpha_{k2}(t)$
are given in terms of (\ref{N22}). The tangle, given by
(\ref{N23}), describes entanglement between the $k$th QD and all
the remaining QDs (denoted by subscript $\tilde k$). The residual
tangles $\Delta^{NM}_{k}$ are then given by
\begin{eqnarray}
  \Delta^{NM}_{\e}(t) &=& \tau_{\e\tilde \e}^{NM}(t)-(M-1)
  [C_{\e\e}^{NM}(t)]^2
\nonumber \\
  &&-(N-M)[C_{\e\g}^{NM}(t)]^2,
\nonumber \\
 \Delta^{NM}_{\g}(t) &=& \tau_{\g\tilde \g}^{NM}(t)-(N-M-1)
  [C_{\g\g}^{NM}(t)]^2
\nonumber \\
  &&-M[C_{\e\g}^{NM}(t)]^2. \label{N24}
\end{eqnarray}
Let us analyze a few examples of the general solutions to describe
generation of $N$-partite entanglement. The evolution, given by
the solution (\ref{N09}), of the initial state
$|\psi^{31}(0)\rangle=|\e\g\g\>$ for three QDs with a single spin
up can explicitly be given by
\begin{eqnarray}
   |\psi^{31}(t)\> &=& \gamma^{31}_{0}(t) |\e \g\g\> +
 \sqrt{2}\gamma^{31}_{1}(t)|\g\>|B\>,
 \label{N25}
\end{eqnarray}
where $\gamma^{31}_{0}(t) =(e^{-i2\kappa t}+2e^{i\kappa t})/3$ and
$\gamma^{31}_{1}(t) =(e^{-i2\kappa t}-e^{i\kappa t})/3$. The state
periodically arrives at, e.g., times $\kappa
t'=(9n+1)\frac{2\pi}9$ into a $W$ state:
\begin{equation}
|\psi ^{31}(t')\rangle =\frac 1{\sqrt{3}}\big( {\rm e}^{i\theta
_0}|\e\g\g\rangle +{\rm e}^{i\theta _1}|\g\e\g\rangle +{\rm e}^{i
\theta _1}|\g\g\e\rangle \big)  \label{N26}
\end{equation}
deviating from the standard $W$ state by the phases $\theta _0
=\arctan \left( \frac{\sqrt{3} s-c }{\sqrt{3}c+s}\right)$ and
$\theta _1 =\arctan \left( \frac{\sqrt{3} s+c
}{\sqrt{3}c-s}\right) -\pi$, where $s=\sin (2\pi /9)$ and $c=\cos
(2\pi /9)$. For the $W$ state one finds from our general solution
that the concurrences are the same and equal to
$C^{31}_{\e\g}(t')=C^{31}_{\g\g}(t')=2/3$, and the tangles are
$\tau^{31}_{\e\tilde\e}(t')=\tau^{31}_{\g\tilde\g}(t')=8/9$, while
the residual tangles are vanishing,
$\Delta^{31}_{\e}(t')=\Delta^{31}_{\g}(t')=0$. Thus, it is clear
the state does not exhibit intrinsic 3-particle entanglement in
agreement with the CKW result \cite{CKW}. For other evolution
times, it is impossible to get higher value of the mutually equal
concurrences (and the tangles) as the W state reaches the upper
bound of $2/N$ for symmetric sharing of entanglement
\cite{Koashi,Dur01}. Nevertheless one of the tangles or the
concurrences can be larger, at a given moment, than those for the
$W$ state. On the other hand, the residual tangles are always
vanishing. These conclusions can be drawn by analyzing the
explicit solutions:
\begin{eqnarray}
  \tau^{31}_{\g\tilde\g}(t) &=& [C^{31}_{\e\g}(t)]^2+[C^{31}_{\g\g}(t)]^2
  = \tau'[7+2\cos(3\kappa t)],
\nonumber \\
  \tau^{31}_{\e\tilde\e}(t) &=& 2[C^{31}_{\e\g}(t)]^2
  = 2 \tau'[5+4\cos(3\kappa t)],
  \label{N27}
\end{eqnarray}
where $\tau'=(16/81)\sin^2(3\kappa t/2)$. Thus, one can observe
that the maximum of concurrence $\max_t C^{31}_{\e\g}(t) =
C^{31}_{\e\g}(t'')$ is 8/9 and the maximum of the tangle $\max_t
\tau^{31}_{\e\tilde\e}(t) = \tau^{31}_{\e\tilde\e}(t'')$ reaches
80/81 for the moments $\kappa t''=\pm 0.565\cdots+n2\pi/3$
($n=0,1,...$). On the other hand, $\max_t C^{31}_{\g\g}(t) =
C^{31}_{\g\g}(t''')=8/9$ and $\max_t \tau^{31}_{\g\tilde\g}(t) =
\tau^{31}_{\g\tilde\g}(t''')=0.9877$ for the moments $\kappa
t'''=(2n+1)\pi/3$. It is seen that these values of the
entanglement measures are much higher than those for the $W$
state.

\begin{table}
\caption{ Maxima of the concurrences $C^{NM}_{\e\e}(t)$,
$C^{NM}_{\e\g}(t)$ and $C^{NM}_{\g\g}(t)$, the tangles
$\tau^{NM}_{\g\tilde\g}(t)$ as well as the residual tangles
$\Delta^{NM}_{\e}(t)$ and $\Delta^{NM}_{\e}(t)$ for
equivalent-neighbor system of $N$ QDs with $M$ of them initially
with spin up. Additionally, $\max_t
\tau^{NM}_{\e\tilde\e}(t)=1.000$ for all presented cases except
$\max_t \tau^{10,2}_{\e\tilde\e}(t)=0.991$. Note that for other
values of $M$, the maxima of the tangles
$\tau^{NM}_{\e\tilde\e}(t)$ also decrease with increasing $N$ if
$N>10$.} \hspace*{-2mm}
\begin{tabular}{|c|c|c|c|c|c|c|c|c|}
\hline N & M  & ${\rm max} C_{\e\e}$  & ${\rm max} C_{\e\g}$  &
${\rm max} C_{\g\g}$ & ${\rm max} \tau_{\g\tilde\g}$ & ${\rm max}
\Delta_{\e}$
& ${\rm max} \Delta_{\g}$ \\
\hline
4 & 2  & 0      &   0.412 &    0     &   1.000 &    0.889 &    0.889 \\
5 & 2  & 0.576  &   0.338 &    0.347 &   1.000 &    0.967 &    0.922 \\
6 & 2  & 0.372  &   0.295 &    0.225 &   0.919 &    0.877 &    0.789 \\
7 & 2  & 0.408  &   0.266 &    0.182 &   0.784 &    0.884 &    0.632 \\
8 & 2  & 0.364  &   0.244 &    0.141 &   0.610 &    0.849 &    0.495 \\
9 & 2  & 0.307  &   0.226 &    0.109 &   0.502 &    0.803 &    0.393 \\
10& 2  & 0.253  &   0.211 &    0.084 &   0.401 &    0.755 &    0.311 \\
\hline
 6& 3&     0 & 0.275 &    0 &1.000 &0.960 &0.960\\
 7& 3& 0.129 & 0.240 &0.191 &1.000 &0.995 &0.987\\
 8& 3& 0.111 & 0.216 &0.115 &0.973 &0.950 &0.928\\
 9& 3& 0.236 & 0.199 &0.135 &0.906 &0.955 &0.811\\
10& 3& 0.205 & 0.174 &0.092 &0.643 &0.886 &0.560\\
\hline
8 & 4&     0 & 0.209 &    0 &1.000 &0.980 &0.980\\
9 & 4&  0.001& 0.188 &0.111 &1.000 &0.999 &0.994\\
10& 4&  0.000& 0.173 &0.034 &0.993 &0.983 &0.967\\
11& 4&  0.136& 0.161 &0.097 &0.966 &0.977 &0.898\\
\hline
\end{tabular}
\end{table}

Similarly for $N=4$, the state $|\psi ^{41}(0)\rangle$ evolves at
times $\kappa t'=(2n+1)\frac \pi 4$ into a four-particle
(generalized) $W$ state
\begin{equation}
|\psi ^{41}(t')\rangle \sim\frac 12 \big(|\e\g\g\g\rangle
-|\g\e\g\g\rangle -|\g\g\e\g\rangle -|\g\g\g\e\rangle \big).
\label{N28}
\end{equation}
Due to the symmetry of the system, $W$ states are also generated
for $|\psi ^{32}(t)\rangle$ and $|\psi ^{43}(t)\rangle$. However,
as can be shown semi-analytically and numerically, the states
$|\psi ^{NM}(0)\rangle$ for $N=4,M=2$ as well as for $N>4$ with
any $M$ do not evolve into the exact $W$ states under the
interaction described by Hamiltonian (\ref{N02}). The concurrences
$C^{41}_{\e\g}(t')=C^{41}_{\g\g}(t')$ reach the value 1/2, which
is the upper bound of symmetric sharing of entanglement for four
qubits \cite{Koashi,Dur01}. Additionally, the tangles are found to
be $\tau^{41}_{\e\tilde\e}(t') =\tau^{41}_{\g\tilde\g}(t')=3/4$.
As in the former example, we want to get higher-values of the
tangles and concurrences by resigning from the condition that
$C^{41}_{\e\g}(t')=C^{41}_{\g\g}(t')$. From the general solutions,
one gets
\begin{eqnarray}
  \tau^{41}_{\g\tilde\g}(t) &=& [C^{41}_{\e\g}(t)]^2+2[C^{41}_{\g\g}(t)]^2
  = \tau'[7+\cos(4\kappa t)],
\nonumber \\
  \tau^{41}_{\e\tilde\e}(t) &=& 3[C^{41}_{\e\g}(t)]^2
  =3 \tau'[5+3\cos(4\kappa t)],
  \label{N29}
\end{eqnarray}
where $\tau'=(1/8)\sin^2(2\kappa t)$. Thus, one can find that
maximum of concurrence $C^{41}_{\e\g}(t)$ (equal to 0.577) and
maximum of tangle $\tau^{41}_{\e\tilde\e}(t)$ (equal to 1) are
higher than those for the W state. However, the maximum of the
other concurrence $C^{41}_{\g\g}(t)$ and tangle
$\tau^{41}_{\e\tilde\e}(t)$ is already reached by the W state.
Moreover, the residual tangles are zero during the whole
evolution.

In general, for the equivalent-neighbor system with arbitrary
number $N$ of QDs and only one of them having initially spin
opposite to the others, we find that
\begin{eqnarray}
   \Delta^{N1}_{\e}(t) &=& \Delta^{N1}_{\g}(t) =0,
\label{N30}
\end{eqnarray}
which implies that the genuine multipartite entanglement is not
generated for $M=1$.

For systems with $M>1$, evolution becomes more complicated as seen
in figures 4 and 5. In the present study, the most interesting for
us are the maximum values of the concurrences and tangles as shown
in table I for various numbers $N$ of QDs and their initial states
as described by number $M$. For example, for any $N$ and $M=2$, we
have
\begin{eqnarray}
  |\psi^{N2}(t) \rangle &=&
   \gamma^{N2}_{0}(t) |\e\e\g\g\>
+2\gamma^{N2}_{1}(t) |BB\> \nonumber \\&& +\gamma^{N2}_{2}(t)
|\g\g\e\e\>, \label{N31}
\end{eqnarray}
where $\gamma^{42}_{0,2}(t) = (e^{-i4\kappa t} + 2e^{i2\kappa
t})/6\pm 1/2$ and $\gamma^{42}_{1}(t) = (e^{-i4\kappa t}
-e^{i2\kappa t})/6$. Such states exhibit genuine multipartite
entanglement as clearly shown in table I and figure 4. Let us
analyze explicitly the simplest state among them, i.e., for $N$=4
and $M$=2. For evolution moment $\kappa t'=\pi/2$, the QD system
evolves into the state
\begin{eqnarray}
  |\psi^{N2}(t') \rangle &=& \frac13 (|\e\e\g\g\>+2|B\>|B\>-2|\g\g\e\e\>),
\label{N32}
\end{eqnarray}
which does not exhibit pairwise entanglement,
$C^{42}_{\e\e}(t')=C^{42}_{\e\g}(t')=C^{42}_{\g\g}(t')=0$, but
exhibits the genuine multipartite entanglement, as described by
the residual tangles
$\Delta^{42}_{\e}(t')=\Delta^{42}_{\g}(t')=8/9$. This value is the
highest during the whole evolution of $|\psi^{42}(t) \rangle$.
Nevertheless higher degree of genuine multipartite entanglement
can be observed by increasing number of QDs to $N=5$ for a fixed
$M=2$ or by increasing both $N$ and $M$. A closer look at the data
shown in table I, enables us to draw a conclusion that the highest
degree of genuine multipartite entanglement, as quantified by the
residual tangle, is generated in the discussed model for odd
number $N=2M+1$ of QDs and by initially preparing $M$ of them in
the state with spin up (or spin down).

\section{Teleportation of electron-spin states}

Quantum teleportation of Bennett \etal~\cite{Bennett93} is a
method to transfer (information about) unknown quantum states over
large distances via entangled particles and transmission of some
classical information. Quantum teleportation is not just a curious
effect but a fundamental protocol which enables universal quantum
computation \cite{Gottesman} as any quantum circuit can be
realized using only quantum teleportation and single-qubit
operations.
\begin{figure}
\includegraphics[width=7cm]{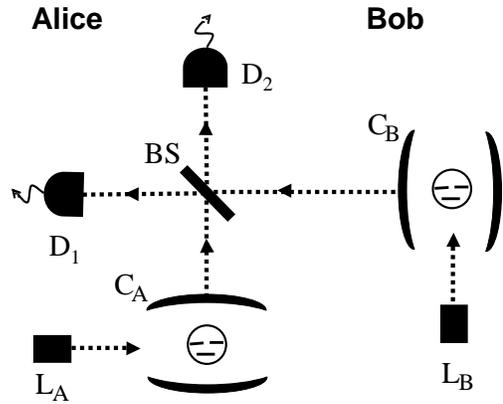}
\caption{Teleportation without insurance of electron-spin states
via cavity decay based on the protocol of Bose \etal~
\protect\cite{Bose}. Key: $C_A,C_B$ - microcavities, $D_A,D_B$ -
photon counters, $L_A,L_B$ - lasers, BS - beam splitter.}
\end{figure}
\begin{figure}
\includegraphics[width=7cm]{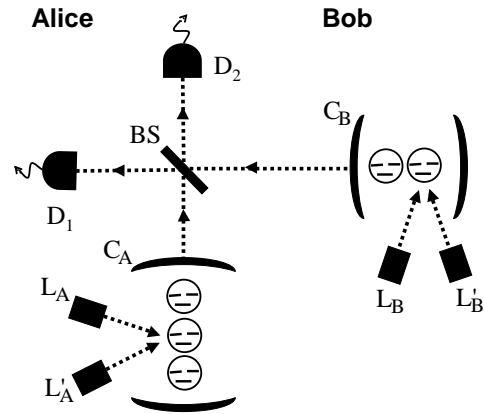}
\caption{Teleportation with insurance of entangled electron-spin
states via cavity decay in our generalized scheme.}
\end{figure}

Here, we describe a protocol for quantum teleportation of
electron-spin states of semiconductor QDs based on the protocol of
Bose \etal~\cite{Bose} (see also \cite{Chimczak1}) for
teleportation of atomic states. The cavity spontaneous photon
leakage plays a crucial role in this teleportation protocol. It is
a well accepted fact that spontaneous decay of excited quantum
systems is a mechanism of their decoherence and therefore usually
plays a destructive role in QIP. However, Bose \etal~have shown
how detection of decay can be used constructively not only for
establishment of entanglement but also for the complete QIP such
as teleportation. This surprising result can be understood by
recalling the fact that a detected decay is a measurement on the
state of the system from which the decay ensues.

Although, the original protocol of Bose \etal~was described for
trapped atoms, it can, as well, be applied to teleportation of
conduction-band electron-spin states of QDs with levels shown in
figure 1. In the original scheme \cite{Bose}, it has been assumed
that $\Delta\omega^{\e}_n=\Delta\omega^{\g}_n$ for $n=1,2$ but one
can resign from this condition. An outline of the modified scheme
is shown in figure 6. The setup consists of two optical cavities:
Alice's $C_A$ and Bob's $C_B$ tuned to the same frequency
$\omega_{\rm cav}$. Each cavity contains a single three-level QD,
which is illuminated within a proper period of time by classical
laser field ($L_A$ or $L_B$). By illuminating the QDs with the
classical laser field of frequency $\omega_n^{L}$, Alice
(designated by subscript $n=A$) and Bob ($n=B$) can drive the
transition between $|\g\rangle_n$ and $|v\rangle_n$. The other
transition between $|\e\rangle_n$ and $|v\rangle_n$ is driven by
the quantized cavity field of frequency $\omega_{\rm cav}$. It is
important to assume that detunings
$\Delta\omega^{\e}_n,\Delta\omega^{\g}_n$ are large enough such
that the lower levels $|v\rangle_n$ can effectively be decoupled
(so neglected) from the evolution of the lower levels. Thus, we
can assume that the quantum information is stored only in two
levels $|\g\rangle_n$ and $|\e\rangle_n$. Both Alice's and Bob's
cavities initially have no photons being described by vacuum state
$|0\rangle_n$, and Bob's QD is initially in state $|\g\rangle$.
Alice does not know her QD state, which is of the form
$|\psi\rangle_A=c |\g\rangle_A + c' |\e\rangle_A$ (with the
unknown coefficients $c$ and $c'$ such that $|c|^2+|c'|^2=1$). The
main task is to teleport the state $|\psi\rangle_A$ to Bob. First,
as a preparation of the state, Alice maps the QD state
$|\psi\rangle_A$ on her cavity mode by illuminating her QD with
the laser $L_A$ for a proper period of time. In the meantime, Bob
illuminates his QD with the laser $L_B$ for another appropriate
time period to generate a QD--cavity-field entangled state
$|\Psi\rangle_B=2^{-1/2} (|\g\rangle_B|1\rangle_B
+i|\e\rangle_B|0\rangle_B)$, where $|1\rangle_B$ and $|0\rangle_B$
stand for the cavity state in vacuum or single-photon state,
respectively. Alice and Bob should synchronize their actions to
finish simultaneously the preparations of their states since
photons are leaking out from both of the cavities. Those photons
are mixed on the 50–-50 beam splitter BS. Cavities are assumed to
be single-sided so that the only leakage of photons occur through
the sides of the cavities facing BS. The next step is the
detection of the photons, when Alice just waits for a finite time
period for click of the photon counter either $D_1$ or $D_2$. This
detection of photons leaking from distinct cavities $C_A$ or $C_B$
constitutes a measurement that enables a transfer of quantum
information from Alice's QD to Bob's QD. The cases, when Alice
registers no clicks or two clicks, are rejected as the failure of
the teleportation. At the post detection stage, Bob applies to the
transferred state a proper phase shift depending on whether
detector $D_1$ or $D_2$ clicked. This step corresponds to the
correcting unitary transformation and completes the teleportation
protocol. It is worth noting that the presented scheme, contrary
to the original Bennett \etal~scheme \cite{Bennett93}, is
probabilistic in the sense that the original state is destroyed
even if the teleportation fails, which is the case when photon
counters do not register one photon. Nevertheless, the scheme can
be modified to a teleportation protocol with insurance by
entangling the initial Alice's QD with a reserve QD also placed in
her cavity $C_A$ to increase the probability of success. In Ref.
\cite{Chimczak2}, we have proposed a generalized scheme, depicted
in figure 7, that allows the teleportation of an entangled state
of two atoms with insurance. Numerical calculations in Refs.
\cite{Chimczak2,Chimczak3} showed that the average probability of
success of our protocol is about $0.94$, while the average
probability of successful teleportation without the insurance does
not exceed $0.5$. Our proposal for teleportation of QD spin states
is a generalization of the scheme for teleportation of
atomic-qubit states via cavity decay
\cite{Bose,Chimczak1,Chimczak2,Chimczak3}. It is a multi-stage
protocol and it is full description would exceed the recommended
number of pages for the proceedings. Thus, the details will be
presented elsewhere \cite{Chimczak4}.

\section{Discussion and conclusion}

Decoherence seriously limits the feasibility of the schemes
especially by comparing decoherence rates in relation to the
gate-operation times. But for simplicity in section III, we
neglected decoherence effects in the analysis of quantum
entanglement in the generalized models of Imamo\v{g}lu
\etal~\cite{Imamoglu}. After Ref.~\cite{Imamoglu}, let us give a
few estimations: The spin decoherence times of the conduction-band
electrons are relatively long and it is reasonable to assume to be
$\sim 1 \mu s$. There are a few mechanisms of decoherence
including spin-orbit coupling and cavity decay. The first
decoherence mechanism is due to the coupling of the
conduction-band electron spins to valence-band holes, which can
result in decoherence of $\sim 1 ns$ and effective decoherence of
$\sim 100ns$. A more deteriorating effect is due to a short cavity
lifetime $\Gamma_{\rm cav}^{-1} \sim 10 ps$, which can result in
the effective decoherence of $\sim 1 ns$ \cite{Imamoglu}. By
contrast, we discussed in section IV how to utilize this cavity
decay in a constructive way for QIP. More details about
decoherence of three-level systems in leaky cavities can be found
in our papers \cite{Chimczak1,Chimczak2,Chimczak3}. Also our
estimations of the size-dependent decoherence of large
semiconductor QDs can be found \cite{Liu03}. Another serious
obstacle to implement the discussed schemes is the requirement of
strong coupling of a QD with a single photon. It seems that
photonic-crystal microcavities could be a good solution, as they
simultaneously exhibit a high quality factor (Q) in excess of
10,000 \cite{Srinivasan} and they are of an ultra-small,
wavelength-scale modal volume. Although our results might appear
highly theoretical, it should be noted that quantum entanglement
of excitons in a single QD~\cite{Chen} and a QD
molecule~\cite{Bayer} has already been observed.

In this communication, we studied possibilities of generation of
maximum pure-state bipartite and multipartite entanglement as well
as mixed-state pairwise entanglement of electron spins in systems
of QDs interacting via a microcavity field within the generalized
models of Imamo\v{g}lu \etal~\cite{Imamoglu}. Conditions for
generation of genuine multipartite entangled states of the
conduction-band electron spins of QDs were discussed based on the
Coffman-Kundu-Wootters inequality. Such problems are important in
the context of possible solid-state implementations of QIP
including quantum teleportation. In particular, we briefly
described a generalized protocol \cite{Chimczak4} of Bose
\etal~\cite{Bose} for teleportation (without and with insurance)
of entangled spin-states of QDs between distant microcavities via
their decay.

\noindent {\bf Acknowledgment.} We acknowledge the support from
the Japan Society for the Promotion of Science within the 21st
Century COE Program and from the Polish KBN (grant No. 1 P03B 064
28).


\end{document}